\documentclass[twocolumn,prX,nofootinbib]{revtex4-1}
\pdfoutput=1
\usepackage{amsmath,amssymb,amsfonts,eucal,mathrsfs,graphicx,tensor}
\usepackage[usenames,dvipsnames]{color}
\usepackage{hyperref}
\DeclareMathAlphabet{\mathpzc}{OT1}{pzc}{m}{it}

\definecolor{brown}{rgb}{0.65,0.16,0.16}

\newcommand{\beql}[1]{\begin{equation}\label{#1}}
\newcommand{\eeq}{\end{equation}}

\newcommand{\fr}{\frac}

\newcommand{\lag}{\mathscr{L}}

\newcommand{\Sg}{\Sigma}

\newcommand{\ed}{\mathrm{d}}

\newcommand{\Dl}{\nabla}
\newcommand{\vecDl}{\vec{\nabla}}

\newcommand{\met}{\mathsf{g}}

\newcommand{\CC}{{\text{\sc cc}}}

\begin{document}
\title{Constructing entanglement wedges for Lifshitz spacetimes with Lifshitz gravity}
\author{Jonathan Cheyne}\email{jacheyne85@googlemail.com}
\author{David Mattingly}\email{david.mattingly@unh.edu}
\affiliation{Department of Physics, University of New Hampshire, Durham, NH 03824, USA}
\begin{abstract}
Holographic relationships between entanglement entropy on the boundary of a spacetime and the area of minimal surfaces in the bulk provide an important entry in the bulk/boundary dictionary. While constructing the necessary causal and entanglement wedges is well understood in asymptotically AdS spacetimes, less is known about the equivalent constructions in spacetimes with different asymptotics.  In particular, recent attempts to construct entanglement and causal wedges for asymptotically Lifshitz solutions in relativistic gravitational theories have proven problematic.  We note a simple observation, that a Lifshitz bulk theory, specifically a covariant formulation of Ho\v{r}ava-Lifshitz gravity coupled to matter, has causal propagation defined by Lifshitz modes.  We use these modes to construct causal and entanglement wedges and compute the geometric entanglement entropy, which in such a construction matches the field theory prescription. 
\end{abstract}

\maketitle

\section{Introduction}

Holographic approaches to field theory and quantum gravity are extremely powerful and at the same time quite limited.  One key limitation is that the best understood arena for holography, namely AdS/CFT, is limited to conformal or near conformal field theories.  Since holographic techniques are powerful, one would like to extend them to non-conformal field theories and systems.  In particular, non-relativistic condensed matter systems, which are not conformal, exhibit many features in principle amenable to holographic calculations if one could extend holography beyond conformal field theory.  Non-relativistic systems often exhibit Lifshitz behavior in certain regimes and therefore developing a Lifshitz holography is a necessary step towards extending holographic techniques to many other important systems~\cite{Ross:2011gu}.

Gravitational backgrounds with Lifshitz symmetries are not solutions to vacuum general relativity, and hence studies of Lifshitz holography have often focused on gravitational models with extra matter fields, for example Einstein-Maxwell-dilaton gravity~\cite{Goldstein:2009cv} or Einstein-Proca gravity~\cite{Taylor:2008tg}. An alternative is provided by modifying gravity itself, and Lifshitz solutions have been found in massive gravity~\cite{AyonBeato:2009nh} and bi-gravity theories as well~\cite{Goya:2014eya}.  In all the above models, however, the Lifshitz nature is a feature of the solutions and not built-in to the theory at a fundamental level.  One modified theory of gravity does, however, intrinsically assume a fundamental Lifshitz symmetry: Ho\v{r}ava-Lifshitz gravity~\cite{Horava:2009uw,Sotiriou:2010wn}, or HL gravity for short.  HL gravity possesses the spacetime manifold structure of general relativity, but additionally equips the manifold with a preferred foliation.  

The requirement of a preferred foliation breaks Lorentz symmetry. As a result there are additional terms allowed in the gravitational action and a modified theory of gravity in both the infrared and ultraviolet.  In the ultraviolet one imposes a Lifshitz symmetry which renders the theory power counting renormalizable without introducing ghosts, unlike what happens in higher curvature relativistic gravity~\cite{Horava:2009uw,Visser:2009fg,Barvinsky:2015kil}. HL gravity comes in various flavors~\cite{Sotiriou:2010wn} - we will be using the generally covariant non-projectable flavor~\cite{Blas:2009qj}.  We make this choice since a) general covariance is a key feature of holographic gravity duals and b) the Lifshitz background spacetime solution we employ is only possible in non-projectable HL gravity.  Due to its likely renormalizability and general covariance, HL gravity serves as both a well-behaved candidate theory of quantum gravity and a possible arena for Lifshitz holography.

Indeed, HL gravity has already proven to be a fertile ground for explorations of Lifshitz holography. Globally Lifshitz solutions exist naturally within HL gravity with a cosmological constant, and have been argued to provide a better gravitational dual for zero temperature Lifshitz field theories~\cite{Griffin:2012qx}.  Asymptotically Lifshitz black holes exist in HL gravity and have a good first law~\cite{Basu:2016vyz}, unlike their asymptotically AdS cousins~\cite{Bhattacharyya:2014kta}.  In this paper we show that HL gravity also may naturally resolve a problem in building causal and entanglement wedges for Lifshitz spacetime that arises in relativistic gravitational theories~\cite{Gentle16}.

Holographic entanglement entropy is an important entry in the holographic dictionary.  In the static case, the entanglement entropy for two disjoint regions in a Lifshitz theory at a moment in time can be computed and matches the holographic calculation~\cite{Nesterov11,Solodukhin10} via a simple extension of the original Ryu-Takayanagi construction~\cite{Ryu06}. Attempts to build a covariant Lifshitz construction, i.e. the equivalent of the Hubeny-Rankangami-Takayanagi construction, have been thwarted by the fact that the entanglement wedges do not naturally reach the boundary of the spacetime~\cite{Gentle16}.  In a relativistic gravitational theory, such as Einstein-Proca or Einstein-Maxwell-dilaton theory, even though Lifshitz spacetime is a solution the wedges are built using relativistic propagation.  This is the essential reason for the obstruction.  

In HL gravity, however, mode propagation is not necessarily relativistic. For example there is an extra massless scalar mode in the theory, and at low energies both it and the usual tensor modes generically propagate with different speeds.   More importantly, at high frequencies mode dispersion relations for both gravitational and matter excitations are non-relativistic, as required by the Lifshitz nature of the ultraviolet theory.  This leads to both a frequency dependent mode speed and arbitrarily fast speed relative to the preferred foliation in the far ultraviolet~\cite{Bhattacharyya:2015gwa}.  High energy Lifshitz behavior has already been argued to be relevant for black hole entropy in  HL gravity~\cite{Berglund:2012fk,Cropp:2013sea}.  We show that one can use causal propagation of these high frequency Lifshitz modes to construct wedges that naturally terminate on the boundary in globally Lifshitz spacetime and reproduce the field theory result for the entanglement entropy.

The paper is organized as follows.  First, we briefly review entanglement entropy in Lifshitz field theories.  We then summarize the Ryu-Takayanagi and Hubeny-Rankangami-Takayanagi constructions in AdS, including the structure of both causal and entanglement wedges, and detail the difficulties encountered in previous attempts to extend both to globally Lifshitz spacetime.  We then introduce HL gravity, show how Lifshitz modes propagating in the bulk in the globally Lifshitz solution propagate to the boundary quite naturally, and construct entanglement and causal wedges that reflect the correct entanglement entropy for Lifshitz field theories. 

\section{Entanglement entropy in Lifshitz field theories}\label{sec:EE}

Before we can begin a discussion of geometric entanglement entropy, we must first note our target: reproducing the entanglement entropy of a Lifshitz field theory.  The entanglement entropy for a region in a Lifshitz field theory can be calculated using the replica trick and  application of the Sommerfeld formula~\cite{Solodukhin10}, just as is demonstrated for other theories in~\cite{Callan94} and ~\cite{Nesterov11}.

Since we are interested in the geometric entanglement entropy the field theory calculational details are of little interest, other than the following result: \textit{the entanglement entropy calculated in this manner is the same in both $z=1$ and $z=2$ cases}~\cite{Solodukhin10}. Here $z$ is the standard Lifshitz exponent of the field theory.  For higher $z$ one must establish the form of the heat kernel which is a distinctly non-trivial problem for $z>2$. We therefore will limit ourselves to $z=2$ for our construction of the geometric entanglement entropy.  Our task is then simple: determine a method for the geometric entropy that yields the same result for both $z=1$ and $z=2$.  

\section{The RT and HRT construction in AdS and Lifshitz space}

\subsection{In AdS}
The original Ryu-Takayanagi conjecture \cite{Ryu06} identified an equivalence between the entanglement entropy of a boundary region and the area of a particular extremal bulk surface in static spacetimes.  More specifically, for any d-dimensional region $A$ on the boundary of $\mathrm{AdS}_{d+2}$ there exists a corresponding d-dimensional bulk surface, $\gamma_A$ such that $\partial A = \partial\gamma_A$ and $\gamma_A$ is a minimal area bulk surface.  They proposed that the entanglement entropy $S_A$ of said boundary region is proportional to $X_{\gamma_A}$, the area of $\gamma_A$, and given by

\begin{equation} \label{eq:entrel}
	S_A=\frac{X_{\gamma_A}}{4G}
\end{equation}
where G is the appropriate gravitational constant.  In order to regularise this result, however, it is necessary, as discussed in \cite{Ryu:2006ef} to impose a cutoff, since otherwise the area diverges at the boundary.

This conjecture is proven in \cite{Fursaev06}, but the minimal area construction is only applicable to a constant time slice, and thus is not covariant.  In \cite{Hubeny07}, several covariant procedures for generating candidate bulk surfaces are considered.  The properties required of these are that they must be covariantly well defined, share a boundary with $A$ (that is, they are anchored at $\partial A$), and must, in the limit of a static spacetime, reduce to the minimal surface proscribed by the procedure previously outlined. Given these requirements, four constructions are discussed in~\cite{Hubeny07}, however we shall focus on two of these, the entanglement wedge, and the causal wedge.

The entanglement wedge is constructed by taking the minimal area bulk surface whose boundary is coincident with that of the boundary region of interest (i.e. occurs at $\partial A$) and constructing light sheets normal to that surface.  Clearly, there will be four sheets, two past and two future.  By selecting that sheet whose cross sectional area converges in each temporal direction, we can define a region bounded by these sheets as the entanglement wedge.  The causal wedge on the other hand, is the region enclosed by the union of the null sheets originating from the past and future boundary points which define the domain of dependence for $A$.  These two wedges (and the corresponding bulk surfaces they define) are not necessarily coincident in all spacetimes, but they do coincide for AdS, as the null sheets involved in the causal wedge intersect at (and normal to) the minimal area bulk surface bounded by $\partial A$  \cite{Hubeny07}.

Attempts thus far to extend these wedge constructions to Lifshitz spacetimes have run into problems, in particular see the discussion in~\cite{Gentle16}.  We outline the essential aspect of the problem below.

\subsection{Failure in Lifshitz spacetime}
Lifshitz spacetimes are similar to AdS spacetime but with anisotropic scaling between temporal and spatial directions.  The line element for 2+1 Lifshitz, which we will concentrate on in this paper for simplicity, is
\beql{eqn:LifLine}
	ds^2=-\frac{R^{2z}}{w^{2z}}dt^2+\frac{R^2}{w^2}dw^2+\frac{R^2}{w^2}dx^2 ,
\eeq
Here $R$ is the Lifshitz scale, $z$ the Lifshitz exponent, $x$ the transverse coordinate, and $w$ the scaled inverse radius coordinate such that asymptotic spatial infinity is at $w=0$.  We denote the timelike Killing vector $d/dt$ by $\tau$ and the transverse Killing vector $d/dx$ by $\chi$.  For $z=1$ \eqref{eqn:LifLine} clearly reduces to the corresponding AdS spacetime.

In a relativistic gravitational theory, causal propagation will be defined by null geodesics of the spacetime.  The corresponding null geodesic equations are 
\begin{eqnarray} 
	t''&=&\frac{2zw't'}{w} \nonumber
	\\ w''&=&z\frac{(t')^2}{w}\left(\frac{R}{w}\right)^{2z-2}+\frac{(w')^2}{w}-\frac{(x')^2}{w} \label{eqn:Lifwdashdash}
	\\ x''&=&\frac{2x'w'}{w}\nonumber
\end{eqnarray}
where $'$ indicates a derivative with respect to some affine parameter $\lambda$. 

For causal wedges to smoothly meet the boundary there must be a causal ray that lies on the boundary given some initial conditions.  In a relativistic theory where the causal rays are null geodesics, we see that imposition of both a null condition ($\mathrm{ds}^2=0$) and a constant radius ($dw=0$) into \eqref{eqn:LifLine} implies
\begin{equation}
	\left(\frac{dx}{dt}\right)^2=\left(\frac{R}{w}\right)^{2z-2}.
\end{equation}
Substitution of this into \eqref{eqn:Lifwdashdash}, shows that for any ray which begins with $w'=0$ in an effort to stay at fixed radius, $w''=0$ if and only if $z=1$, i.e. the AdS case.  For $z>1$, $w''>0$ and thus null rays that start on the boundary are accelerated into the bulk.  This prevents closure of the causal or entanglement wedge in a natural fashion~\cite{Gentle16}.  In the case of the entanglement wedge, the only null geodesic orthogonal to a bulk surface which reaches the boundary is the entirely radially directed one central to the surface, all others are accelerated inwards and give rise to caustics, failing to close the wedge.  For the causal wedge, as shown in Fig. 1, no null geodesics of constant radius exist in a Lifshitz spacetime.  Therefore we cannot generate any wedge which asymptotes to the boundary domain of dependence.  We now turn to how working with Lifshitz modes in HL gravity solves this issue.
\begin{figure}[htb]
	\begin{center}
		\includegraphics[width=0.4\textwidth]{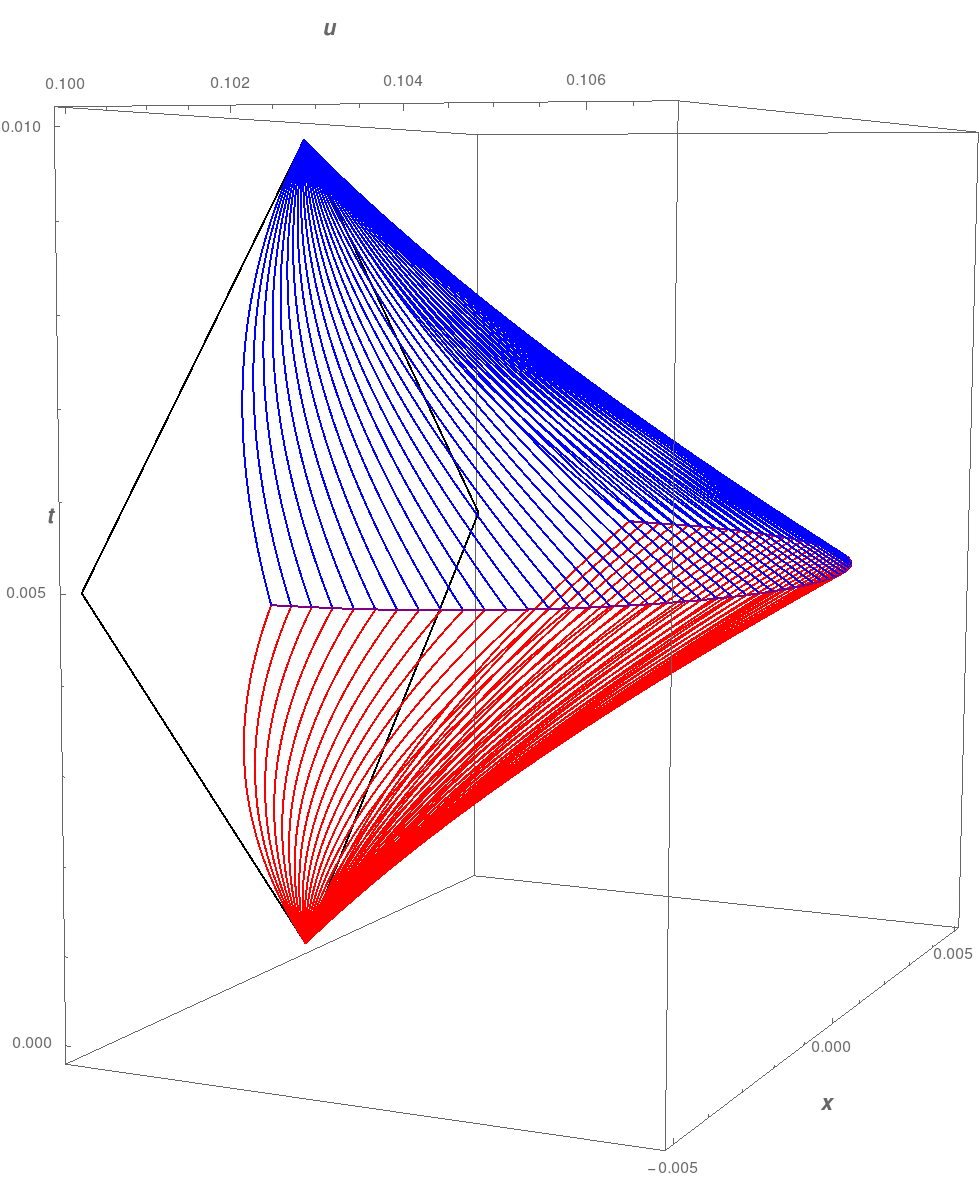}
		\caption{The causal wedge as generated by null geodesics of the Lifshitz spacetime originating at $x=0$, $w=\epsilon$, $t=0$ and by past directed null geodesics originating at $x=0$, $w=\epsilon$, $t=0.01$. For this particular example, z=2,$\epsilon=0.1$.}
	\end{center} \label{fig:geowedge}
\end{figure}

\section{Closing wedges in HL gravity} \label{sec:Horava}
\subsection{HL Action and global Lifshitz solution}
HL gravity has a number of formulations, both covariant and non-covariant.  As mentioned in the introduction, we use the covariant low-energy formulation~\cite{Blas:2009qj} which is closely related~\cite{Jacobson:2010mx} to Einstein-{\ae}ther theory~\cite{Jacobson:2000xp}.  In this formulation the foliation is dynamical and the leaves of the foliation are labelled by a scalar field $T$, called the~\emph{khronon}, which always admits a non-zero timelike gradient everywhere on-shell.  From $T$ one can construct a unit-timelike hypersurface orthogonal one-form $u_a$, called the~\emph{{\ae}ther}, such that
\beql{ae:HSO:norm}
u_a = -N\Dl_a T, \qquad \met^{a b}u_a u_b = -1~,
\eeq
where the function $N$ is solved for via the unit norm constraint as follows 
\beql{eq:lapse}
N^{-2} = -\met^{a b}(\Dl_a T)(\Dl_b T)~.
\eeq
Khronon reparameterization $T\rightarrow \tilde{T}(T)$, where $\tilde{T}$ is a monotonic function of $T$, is a symmetry of the theory.  Therefore the action can be only a function of the {\ae}ther field, which is reparameterization invariant.  Neglecting boundary terms, the low energy action in 2+1 dimensions is 
\beql{action}
S = \frac{1}{16\pi G_{\ae}}\int\ed^3x\sqrt{-\met}(-2\Lambda_{\CC} + R + \lag)
\eeq
where $\Lambda_{\CC}$ is the (negative) cosmological constant, $R$ is the Ricci scalar, and $\lag$ is the khronon's Lagrangian  given by
\beql{lag:ae}
\lag = -\tensor{Z}{^{a b}_{c d}}(\Dl_a u^c)(\Dl_b u^d)~.
\eeq
The tensor $\tensor{Z}{^{a c}_{c d}}$ is given by
\beql{def:Zabcd}
\tensor{Z}{^{a b}_{c d}} = c_1\met^{a b}\met_{c d} + c_2\tensor{\delta}{^a_c}\tensor{\delta}{^b_d} + c_3\tensor{\delta}{^a_d}\tensor{\delta}{^b_c} - c_4u^au^b\met_{c d}~,
\eeq
where $c_1$, $c_2$, $c_3$, $c_4$ are coupling constants. Variation with respect to the metric and khronon then gives the equations of motion.

Clearly a solution in HL gravity is composed of both metric and {\ae}ther profiles.  As shown in ~\cite{Griffin:2012qx, Basu:2016vyz}, the global Lifshitz spacetime~\eqref{eqn:LifLine} introduced in~\cite{Kachru:2008yh} is a solution of the field equations that arise from varying the action~\eqref{action} in conjunction with the aether profile
\beql{u_a:global:EF}
u_a =-\left(\frac{R}{w}\right)^z dt.
\eeq
In other words, the aether vector $u^a$ is the unit time-like vector aligned with $\tau$ everywhere.   $\Lambda_{\CC}$ and $c_1,c_4$ fully determine the parameters in the Lifshitz solution via
\beql{global:CC-c14}
\Lambda_{\CC} = -\frac{z(z + 1)}{2R^2}~, \qquad c_1+c_4 = \frac{z - 1}{z}~.
\eeq

\subsection{Lifshitz modes in Lifshitz spacetime}
\subsubsection{Action and dispersion} \label{subsubsec:scalar}
The high energy Lifshitz symmetry inherent in HL gravity will, of course, feed into the matter sector as well either directly or via loop corrections involving gravitons.  Propagation of matter excitations can therefore be expected to have the same behavior as gravitational excitations: relativistic at low energies with a constant speed (generically not equal to the speed of light)~\cite{Jacobson:2004ts} and non-relativistic at high energies with the dispersion relation controlled by the Lifshitz symmetry.  When constructing causal wedges one should consider all possible modes which means the non-relativistic behavior of high energy modes must be taken into account.  Indeed, the non-relativistic modes are precisely what allows us to remedy the problems encountered in~\cite{Gentle16} when constructing relativistic causal wedges.  

With the {\ae}ther field $u_a$ in hand, we can construct covariant Lagrangians for matter that yield non-relativistic dispersion at high energies. In principle the Lifshitz scaling for the spacetime solution and the matter fields can be different.  We do not consider this case here as the simplest case - with $z$ universal between the Lifshitz background and the matter field - allows us to neatly construct causal wedges.  We also specialize to $z=2$ for both the background solution and the non-relativistic, Lifshitz dispersion since, as previously mentioned in~\ref{sec:EE} that is the only $z$ for which the field theory calculation is fully under control~\cite{Solodukhin10} and we are interested in establishing a match between the geometric entanglement entropy and the field theory calculation.  We stress however that in principle there is no reason why higher $z$ cannot be implemented on the geometric side of the duality. 

Specializing to $z=2$ we take our matter to be a real scalar field with Lagrangian 
\beql{lag:ae-phi:O4}
\lag = -\fr{s_\phi^2}{2}\met_{(\phi)}^{a b}(\Dl_a\phi)(\Dl_b\phi) - \fr{(\vecDl^2\phi)^2}{2k_0^2}~,
\eeq
where $\met_{(\phi)}^{a b} = \met^{a b} -  (s_\phi^{-2} - 1)u^a u^b$,$\vecDl_a$ is the projected (spatial) covariant derivative on a leaf $\Sg_T$, and $s_\phi$ is the low energy speed of the $\phi$-excitations. Since we are interested in Lifshitz modes with $k \gg k_0$ the exact value of $s_\phi$ is irrelevant and we choose it to be one.  Signs are such that all modes are propagating in flat space.  The sign of the $k_0$ term is further chosen such that excitations with momenta greater than $k_0$ always have positive frequency, reflecting the ultraviolet complete Lifshitz nature of HL gravity.

To proceed further in determining the dispersion and propagation of modes we note that since we are interested in high frequency modes we can employ the geometric optics approximation.  A (scalar) mode in the geometric optics approximation is given by
\begin{equation}
\phi(x^a)=A(x^a)e^{i \Phi(x^a)}
\end{equation}
where the amplitude $A(x^a)$ is taken to be slowly varying. The four momentum $k_a$ is given by $k_a=\nabla_a \Phi$ and is also assumed to be slowly varying ($\partial_w k_a \ll k_a$).  Since there are two Killing vectors and two conserved energy/momenta we can rewrite the phase $\Phi$ as,
\begin{equation} \label{eqn:phase}
\Phi=-\Omega t + P x + \int^w k(w') dw'
\end{equation}
where $\Omega$, $P$, and $k$ are the conserved Killing energy, transverse momentum, and longitudinal momentum.  

Substituting ~\eqref{eqn:phase} into the equations of motion generated by varying~\eqref{lag:ae-phi:O4} and evaluating on the Lifshitz background specified by~\eqref{eqn:LifLine} and ~\eqref{u_a:global:EF} yields the following equation for $k$

\begin{eqnarray} \label{eqn:keom}
w^3 \left(2 k_0^2 \Omega ^2+2 i k'''\right) +k \left(8 w^3 k''+4 i k_0^2 R^2\right) \nonumber \\
-i k' \left(2 k_0^2 R^2 w+4 P^2 w^3\right)+6 w^3 k'^2 \nonumber \\
=k^2 \left(12 i w^3 k'+2 k_0^2 R^2 w+4 P^2 w^3\right) \nonumber \\
+2 w^3 k^4+2 k_0^2 R^2 \text{P}^2 w+2 P^4 w^3
\end{eqnarray}
where a prime denotes a derivative with respect to $w$.  In the geometric optics limit, one assumes the momentum $k(w)$ varies slowly relative to the phase and therefore sets $k'\approx k''\approx k'''\approx 0$ above. Since $w \rightarrow 0$ is the boundary at infinity, eventually this approximation must break down for a mode with finite Killing energy $\Omega$.  We remedy this by utilizing the cutoff already necessary for regularizing the entropy, which we call $w_0$. For any choice of $w_0$ there are Killing energies where the geometric approximation still holds.  When we eventually take $w_0 \rightarrow 0$ to reach the boundary at infinity, we must take a double limit $w \rightarrow 0, \Omega \rightarrow \infty$ such that the geometric approximation still holds.  This poses no problem in principle as there is no upper bound on $\Omega$.  However, we caution the reader that a cutoff imposed in this manner and its corresponding implication for the necessary Killing frequencies considered must eventually be reconciled with any cutoff employed in the calculation of the entanglement entropy from the field theory side.  We will not pursue this compatibility here, however, as we are simply concentrating on a geometric construction.

Applying the geometric optics limit the real part of~\eqref{eqn:keom} leads to the dispersion relation
\beql{eq:disp}
 \Omega^2= \frac{R^2}{w^2}  ( k^2 +  P^2)  + \frac{1}{k_0^2}  (P^2 +  k^2)^2
\eeq
which clearly shows quadratic (relativistic) behavior as $k \ll k_0$ and Lifshitz behavior at $k \gg k_0$.\footnote{The reader may be concerned that since we set $s_\phi=1$ we are missing the possibility of constructing causal wedges using relativistic modes that simply travel faster than the usual speed of light.  One can quickly show that this is not possible by evaluating the corresponding line element~\eqref{eqn:LifLine} and geodesic equation~\eqref{eqn:Lifwdashdash}.} The key feature of the dispersion is that while the relativistic part depends on $w$, the non-relativistic part does not, which leads to natural causal wedges, as we now show.

\subsubsection{Propagation of Lifshitz modes}
Lifshitz modes are those that satisfy
\beql{eqn:Lifcond}
P^2 + k^2 \gg \frac{k_0^2 R^2}{w_0^2}
\eeq
and hence have the $z=2$ Lifshitz dispersion
\beql{eq:dispLif}
\Omega^2=  \frac{1}{k_0^2}  (P^2 +  k^2)^2.
\eeq
Solving the dispersion for $k$ yields 
\beql{eqn:kLif}
k= \sqrt{k_0 \Omega - P^2}.
\eeq
As required from the geometric optics approximation, $k$ is a conserved quantity as a Lifshitz mode propagates in a Lifshitz spacetime.  The group velocity $\vec{v}_g=(2k/k_0,2P/k_0)$ is therefore constant and has the familiar non-relativistic form of a particle in free space. As expected, the net result is that a Lifshitz excitation of high enough total energy $\Omega \gg k_0^2 R^4/w_0^4$ in Lifshitz spacetime propagates like a free non-relativistic particle of mass $k_0/2$ and fixed coordinate speed $v=2 \sqrt{\Omega/k_0}$.

\subsection{Wedges and Ryu-Takayanagi entanglement}
We now wish to construct the appropriate wedges and calculate the corresponding Ryu-Takayangi expression for entanglement entropy.  In a relativistic field theory, \textit{which} mode one picks to define the causal wedge is irrelevant as they all travel at the same speed.  In the Lifshitz case each mode has a different group velocity and hence a different causal wedge.  It is therefore impossible to define ``the'' causal wedge without putting a high energy cutoff on $\Omega$ to define the maximum speed excitation.  Such a construction may be possible in a natural way by identifying $w_0$ and the highest allowed Killing energy  $\Omega_0$ holographically, but for calculation of entanglement entropy via Ryu-Takayanagi such a detailed construction is not even necessary.  As we shall see below, the causal and entanglement wedges again coincide and every choice of $\Omega_0$ gives the same value for the entanglement entropy for a given $w_0$.

To see this, consider a region $A$ of coordinate length $2 L_B$ on the boundary at $w_0$ with endpoints $\partial A=\pm L_B$ and modes with Killing frequency $\Omega_0$. The intersection of the causal wedge with the boundary at $w_0$ is generated by tracing signals with $k=0$ emitted from $(0,w_0,\pm L_B)$ forward and backwards in time.  Such signals meet at the points $t_\pm=(\pm L_B \sqrt{k_0/4 \Omega_0} , w_0, 0)$ and their corresponding rays define the intersection of the causal wedge with the boundary at $w=w_0$. The causal wedge in the bulk is generated by propagating signals with energy $\Omega_0$ and bulk longitudinal momentum $-k_0 \Omega_0 <P < k_0 \Omega_0$ that satisfy the dispersion~\eqref{eq:dispLif}.  Since the coordinate speed is fixed by $\Omega_0$ the light cones intersect along a semicircle in the $t=0$ plane of coordinate distance $L_B$ from $(0,w_0,0)$.  This semicircle is \textit{independent} of $\Omega_0$, as $\Omega_0$ only changes the ``height'' of the wedges, i.e. the value of $t_\pm$, but not the curve of intersection of the past/future light cones from $t_\pm$ as that depends solely on $L_B$. 

We now note that the causal wedge coincides with the entanglement wedge, similar to the AdS case.  The null rays of the AdS metric
\beql{eqn:AdSLine}
ds^2=\frac{R^2}{w^2}\left(-dt^2+dw^2+dx^2\right) 
\eeq
clearly also are generated by rays that have a fixed coordinate speed.  Since the only thing that changes is the height, the same argument for coincidence holds.  

As a result of the above the chosen value of $\Omega_0$ is irrelevant for determining the geometric entanglement entropy, just as in the relativistic case.   The only difference is that given some boundary region $A$ the wedge height in the Lifshitz case depends on $\Omega_0$.  The spacelike intersection of the wedges $\gamma_A$ with boundary $\partial A$ is identical, however, as shown in Fig. 2. We caution the reader that this feature is only true in the globally Lifshitz case - in asymptotically Lifshitz spacetimes one cannot necessarily expect this to hold.

The geometric entanglement entropy is now trivial to calculate: it proceeds exactly as it does in the relativistic case.  Since the geometry of the constant $t$ hypersurfaces is identical to the AdS case (only the metric in the time coordinate changes between the Lifshitz and AdS cases), the proper length of $\gamma_A$ remains unchanged. Therefore the numerical result for the geometric entropy is given again by~\eqref{eq:entrel}. This matches the result from the field theory side, where the entanglement entropy for a $z=2$ field theory is equal to that for $z=1$.  
\begin{figure}[htb]
	\begin{center}
		\includegraphics[width=0.4\textwidth]{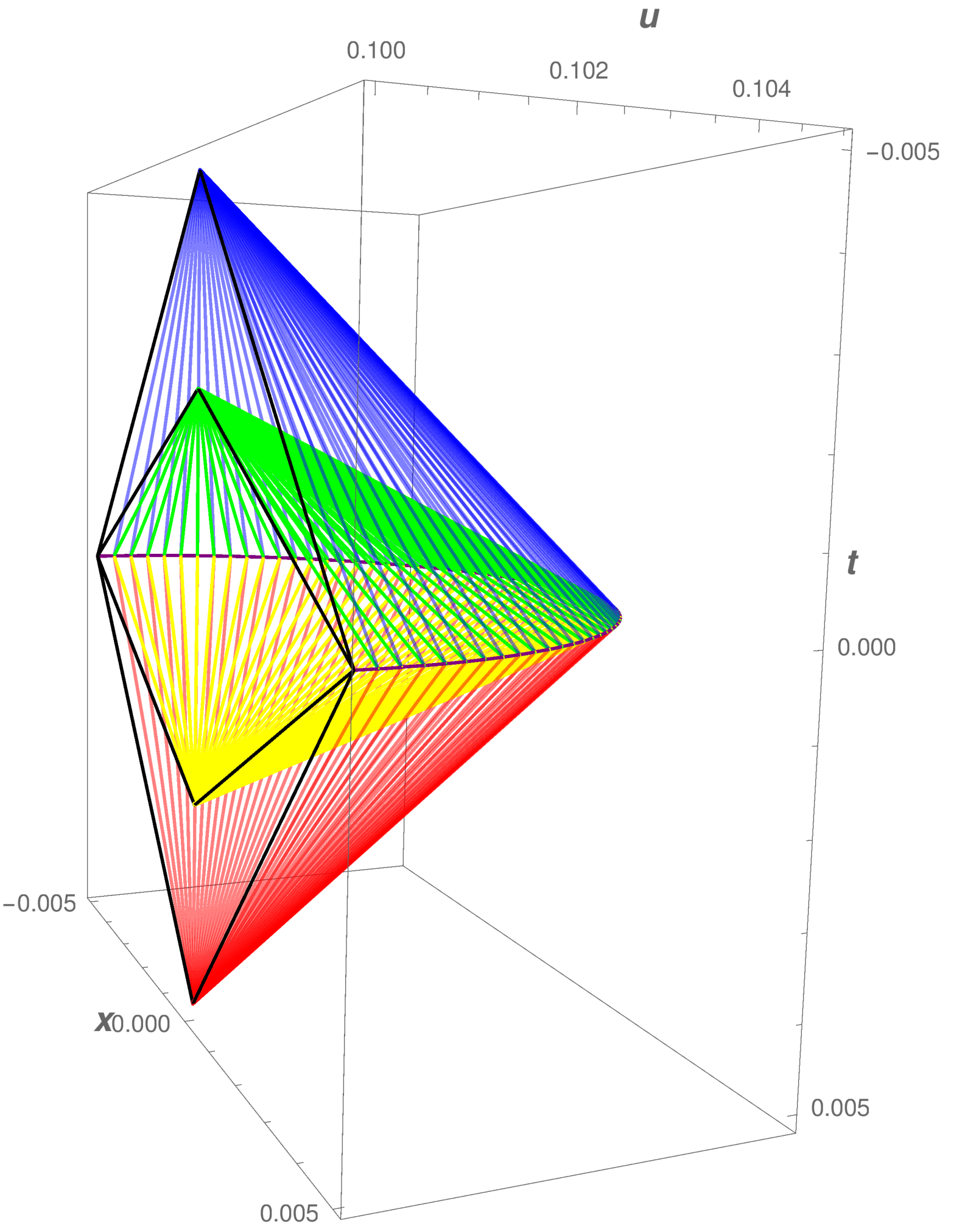}
		\caption{The causal wedges as generated by two distinct Lifshitz modes with the same bondary region $A$ but different Killing frequencies $\Omega_0$. In this case $\Omega_0$ for the shorter wedge is twice that of the other.}
	\end{center} \label{fig:twowedges}
\end{figure}

\section{Conclusion}
Constructing causal/entanglement wedges in a Lifshitz geometry using relativistic causality has proven problematic, which made a construction of the Ryu-Takayanagi geometric entropy also difficult, even though the field theory calculations give identical results.  We have shown that if one uses a Lifshitz gravitational field theory, in particular HL gravity, that supports both a Lifshitz spacetime and a non-relativistic causal structure at high energies a very natural construction emerges that in principle reproduces the field theory calculation.  While there are numerous checks to be done to see if this proposal is actually viable, HL theory again seems to be a very natural candidate bulk theory for implementing non-relativistic holography.

We finally note that the use of Lifshitz modes for the causal and entanglement wedges provides a possible route to relating black hole entropy in HL theory to geometric entanglement entropy, similar to how it can be done in AdS/CFT~\cite{Emparan:2006ni} via the HRT conjecture.  The appropriate causal horizons in HL theory are universal horizons, not Killing horizons, and black hole thermodynamics appears to apply to these horizons instead (although there are still unanswered questions).  Of particular note is that only for very high frequency Lifshitz modes is a thermal spectrum dictated by the surface gravity at the universal horizon expected to be seen at infinity - lower energy modes scatter heavily off the Killing horizon as they propagate outwards.  Since the modes used in this work are precisely the high energy Lifshitz modes, the Killing horizon is again irrelevant, and there is a natural geometric entropy, it is reasonable to expect that, just as in relativistic theories, one can rewrite horizon entropy as geometric entanglement entropy.  We leave this question for future work.

\begin{acknowledgements}
The authors thank Jishnu Bhattacharyya for useful conversations and suggestions.	
\end{acknowledgements}
	
\bibliographystyle{unsrt}

\begin{thebibliography}{99}

\bibitem{Ross:2011gu} 
S.~F.~Ross,
Class.\ Quant.\ Grav.\  {\bf 28}, 215019 (2011)
doi:10.1088/0264-9381/28/21/215019
[arXiv:1107.4451 [hep-th]].
	
	\bibitem{Goldstein:2009cv} 
	K.~Goldstein, S.~Kachru, S.~Prakash and S.~P.~Trivedi,
	JHEP {\bf 1008}, 078 (2010)
	doi:10.1007/JHEP08(2010)078
	[arXiv:0911.3586 [hep-th]].
	
	\bibitem{Taylor:2008tg} 
	M.~Taylor,
	arXiv:0812.0530 [hep-th].
	
	\bibitem{AyonBeato:2009nh} 
	E.~Ayon-Beato, A.~Garbarz, G.~Giribet and M.~Hassaine,
	Phys.\ Rev.\ D {\bf 80}, 104029 (2009)
	doi:10.1103/PhysRevD.80.104029
	[arXiv:0909.1347 [hep-th]].
	
	\bibitem{Goya:2014eya} 
	A.~F.~Goya,
	JHEP {\bf 1409}, 132 (2014)
	doi:10.1007/JHEP09(2014)132
	[arXiv:1406.4771 [hep-th]].
	
	\bibitem{Horava:2009uw} 
	P.~Ho\v{r}ava,
	Phys.\ Rev.\ D {\bf 79}, 084008 (2009)
	doi:10.1103/PhysRevD.79.084008
	[arXiv:0901.3775 [hep-th]].
	
	\bibitem{Sotiriou:2010wn} 
	T.~P.~Sotiriou,
	J.\ Phys.\ Conf.\ Ser.\  {\bf 283}, 012034 (2011)
	doi:10.1088/1742-6596/283/1/012034
	[arXiv:1010.3218 [hep-th]].
		
	\bibitem{Visser:2009fg} 
	M.~Visser,
	Phys.\ Rev.\ D {\bf 80}, 025011 (2009)
	doi:10.1103/PhysRevD.80.025011
	[arXiv:0902.0590 [hep-th]].
	
	\bibitem{Barvinsky:2015kil} 
	A.~O.~Barvinsky, D.~Blas, M.~Herrero-Valea, S.~M.~Sibiryakov and C.~F.~Steinwachs,
	Phys.\ Rev.\ D {\bf 93}, no. 6, 064022 (2016)
	doi:10.1103/PhysRevD.93.064022
	[arXiv:1512.02250 [hep-th]].		
	
	\bibitem{Blas:2009qj} 
	D.~Blas, O.~Pujolas and S.~Sibiryakov,
	Phys.\ Rev.\ Lett.\  {\bf 104}, 181302 (2010)
	doi:10.1103/PhysRevLett.104.181302
	[arXiv:0909.3525 [hep-th]].

	\bibitem{Griffin:2012qx} 
	T.~Griffin, P.~Hořava and C.~M.~Melby-Thompson,
	Phys.\ Rev.\ Lett.\  {\bf 110}, no. 8, 081602 (2013)
	doi:10.1103/PhysRevLett.110.081602
	[arXiv:1211.4872 [hep-th]].
	
	\bibitem{Basu:2016vyz} 
	S.~Basu, J.~Bhattacharyya, D.~Mattingly and M.~Roberson,
	Phys.\ Rev.\ D {\bf 93}, no. 6, 064072 (2016)
	doi:10.1103/PhysRevD.93.064072
	[arXiv:1601.03274 [hep-th]].
		
	\bibitem{Bhattacharyya:2014kta} 
	J.~Bhattacharyya and D.~Mattingly,
	Int.\ J.\ Mod.\ Phys.\ D {\bf 23}, no. 13, 1443005 (2014)
	doi:10.1142/S0218271814430056
	[arXiv:1408.6479 [hep-th]].
						
	\bibitem{Gentle16} 
	S.~A.~Gentle and C.~Keeler,
	JHEP {\bf 1603}, 195 (2016)
	doi:10.1007/JHEP03(2016)195
	[arXiv:1512.04538 [hep-th]].

	\bibitem{Nesterov11}D.~Nesterov and S.~N.~Solodukhin,
	Nucl.\ Phys.\ B {\bf 842}, 141 (2011)
	doi:10.1016/j.nuclphysb.2010.08.006
	[arXiv:1007.1246 [hep-th]].
	
	\bibitem{Solodukhin10}S.~N.~Solodukhin,
	JHEP {\bf 1004}, 101 (2010)
	doi:10.1007/JHEP04(2010)101
	[arXiv:0909.0277 [hep-th]].
	
	\bibitem{Ryu06} 
	S.~Ryu and T.~Takayanagi,
	Phys.\ Rev.\ Lett.\  {\bf 96}, 181602 (2006)
	doi:10.1103/PhysRevLett.96.181602
	[hep-th/0603001].
		
	\bibitem{Bhattacharyya:2015gwa} 
	J.~Bhattacharyya, M.~Colombo and T.~P.~Sotiriou,
	Class.\ Quant.\ Grav.\  {\bf 33}, no. 23, 235003 (2016)
	doi:10.1088/0264-9381/33/23/235003
	[arXiv:1509.01558 [gr-qc]].
	
	\bibitem{Berglund:2012fk} 
	P.~Berglund, J.~Bhattacharyya and D.~Mattingly,
	Phys.\ Rev.\ Lett.\  {\bf 110}, no. 7, 071301 (2013)
	doi:10.1103/PhysRevLett.110.071301
	[arXiv:1210.4940 [hep-th]].
	
	\bibitem{Cropp:2013sea} 
	B.~Cropp, S.~Liberati, A.~Mohd and M.~Visser,
	Phys.\ Rev.\ D {\bf 89}, no. 6, 064061 (2014)
	doi:10.1103/PhysRevD.89.064061
	[arXiv:1312.0405 [gr-qc]].
	
	
	\bibitem{Callan94}C.~G.~Callan, Jr. and F.~Wilczek,
	Phys.\ Lett.\ B {\bf 333}, 55 (1994)
	doi:10.1016/0370-2693(94)91007-3
	[hep-th/9401072].

	\bibitem{Ryu:2006ef} 
	S.~Ryu and T.~Takayanagi,
	JHEP {\bf 0608}, 045 (2006)
	doi:10.1088/1126-6708/2006/08/045
	[hep-th/0605073].

	\bibitem{Fursaev06} 
	D.~V.~Fursaev,
	JHEP {\bf 0609}, 018 (2006)
	doi:10.1088/1126-6708/2006/09/018
	[hep-th/0606184].

	\bibitem{Hubeny07} 
	V.~E.~Hubeny, M.~Rangamani and T.~Takayanagi,
	JHEP {\bf 0707}, 062 (2007)
	doi:10.1088/1126-6708/2007/07/062
	[arXiv:0705.0016 [hep-th]].

\bibitem{Jacobson:2010mx} 
T.~Jacobson,
Phys.\ Rev.\ D {\bf 81}, 101502 (2010)
Erratum: [Phys.\ Rev.\ D {\bf 82}, 129901 (2010)]
doi:10.1103/PhysRevD.82.129901, 10.1103/PhysRevD.81.101502
[arXiv:1001.4823 [hep-th]].

\bibitem{Jacobson:2000xp} 
T.~Jacobson and D.~Mattingly,
Phys.\ Rev.\ D {\bf 64}, 024028 (2001)
doi:10.1103/PhysRevD.64.024028
[gr-qc/0007031].

\bibitem{Kachru:2008yh} 
S.~Kachru, X.~Liu and M.~Mulligan,
Phys.\ Rev.\ D {\bf 78}, 106005 (2008)
doi:10.1103/PhysRevD.78.106005
[arXiv:0808.1725 [hep-th]].

\bibitem{Jacobson:2004ts} 
T.~Jacobson and D.~Mattingly,
Phys.\ Rev.\ D {\bf 70}, 024003 (2004)
doi:10.1103/PhysRevD.70.024003
[gr-qc/0402005].

\bibitem{Emparan:2006ni} 
R.~Emparan,
JHEP {\bf 0606}, 012 (2006)
doi:10.1088/1126-6708/2006/06/012
[hep-th/0603081].


\end{thebibliography}

\end{document}